%% file: main.tex
\documentclass[prl,reprint,superscriptaddress,amsmath,amssymb,aps,longbibliography]{revtex4-2}

\input{packages}
\input{macros}

\begin{document}

\title{Resisting high-energy impact events through gap engineering in superconducting qubit arrays}

\date{\today}
\input{authors}

\input{content/0_abstract} 

\maketitle

%%%%%%%%%%%%%%%%%%%%%%%%%%%%%%%%%%%%%%%%%%%%%%%%%%%%%%%%%%%%%%%%%%%%%%%%%%%%%
% Main Paper Content
%%%%%%%%%%%%%%%%%%%%%%%%%%%%%%%%%%%%%%%%%%%%%%%%%%%%%%%%%%%%%%%%%%%%%%%%%%%%%
\input{content/1_introduction}
\input{content/2_experiment}
\input{content/3_correlated_sampling}
\input{content/4_illumination}
\input{content/5_conclusion}
\input{content/6_declarations}

%\printbibliography
\bibliography{bib.bib, bib_new_entries.bib}

\end{document}

\title{Supplementary Materials for "Resisting quasiparticle bursts through gap-engineering"}
\date{\today}
\input{authors}

\maketitle

\renewcommand\thefigure{S\arabic{figure}} 
\renewcommand\thetable{S\arabic{table}} 

%%%%%%%%%%%%%%%%%%%%%%%%%%%%%%%%%%%%%%%%%%%%%%%%%%%%%%%%%%%%%%%%%%%%%%%%%%%%%
% Supp Paper Content
%%%%%%%%%%%%%%%%%%%%%%%%%%%%%%%%%%%%%%%%%%%%%%%%%%%%%%%%%%%%%%%%%%%%%%%%%%%%%

\input{content/S1_device}

\input{content/S2_rrecs_series}
\input{content/S3_illumination}
\input{content/S4_qp_model}

%\printbibliography
\bibliography{bib.bib, bib_new_entries.bib}

%% file: packages.tex
\usepackage{natbib}
\usepackage{amsmath}
\usepackage{amssymb}
\usepackage{amsthm}
\usepackage{amsfonts}
\usepackage[caption=false]{subfig}
\usepackage[colorlinks]{hyperref}
\usepackage[all]{hypcap}
\usepackage{tikz}
\usepackage{relsize}
\usepackage{color,soul}
\usepackage[utf8]{inputenc}
\usepackage{capt-of}
\usepackage{mathtools}
\usepackage{float}
\usepackage[section]{placeins}
\usepackage{listings}
\usepackage{svg}
\usepackage{xifthen}
\usepackage{multirow}
\usepackage{makecell}
\usepackage{siunitx}

\usepackage[T1]{fontenc}  % Without this, underscores lost when copy-pasting output!
\usetikzlibrary{decorations.pathreplacing}
\usepackage{braket}

%% file: macros.tex
\theoremstyle{definition}

\theoremstyle{definition}

\theoremstyle{definition}

\newcommand{\eq}[1]{\hyperref[eq:#1]{Eq.~\ref*{eq:#1}}}
\renewcommand{\sec}[1]{\hyperref[sec:#1]{Section~\ref*{sec:#1}}}
\newcommand{\fig}[2][]{\hyperref[fig:#2]{Fig.~\ref*{fig:#2}\ifx$#1$\else(#1)\fi}}
\newcommand{\tbl}[1]{\hyperref[tbl:#1]{Table~\ref*{tbl:#1}}}

\newcommand{\goog}{\affiliation{Google Quantum AI, Santa Barbara, CA 93117, USA}}

\newcommand{\yale}{\affiliation{Applied Physics Dept., Yale University, New Haven, CT 06520, USA}}

\newcommand{\us}{$\mu$s }

%% file: authors.tex
\author{Matt McEwen}
\email{mmcewen@google.com}
\thanks{Authors contributed equally}

\author{Kevin C.~Miao}
\thanks{Authors contributed equally}

% strong numbers
\author{Juan Atalaya}

% TLS bro
\author{Alexander Bilmes}

% proposed and made the device
\author{Alex Crook}
\author{Jenna Bovaird}
\author{John~Mark Kreikebaum}

% built chaco, the light stuff, software
\author{Nicholas Zobrist}
\author{Evan Jeffrey}
\author{Bicheng Ying}

% worked on adjacent gap-eng devices / data-taking
\author{Andreas Bengtsson}
\author{Hung-Shen Chang}
\author{Andrew Dunsworth}

% worked on the broader T1 problem
\author{Julian Kelly}
\author{Yaxing Zhang}
\author{Ebrahim Forati}

% developed fab stuff
\author{Rajeev Acharya}
\author{Justin Iveland}
\author{Wayne Liu}
\author{Seon Kim}
\author{Brian Burkett}

% managers
\author{Anthony Megrant}
\author{Yu Chen}
\author{Charles Neill}
\author{Daniel Sank}
\goog

\author{Michel Devoret}
\goog
\yale

\author{Alex Opremcak}
\thanks{Authors contributed equally}
\goog

%% file: content/0_abstract.tex
\begin{abstract}
Quantum error correction (QEC) provides a practical path to fault-tolerant quantum computing through scaling to large qubit numbers, assuming that physical errors are sufficiently uncorrelated in time and space.
In superconducting qubit arrays, high-energy impact events can produce correlated errors, violating this key assumption.
Following such an event, phonons with energy above the superconducting gap propagate throughout the device substrate, which in turn generate a temporary surge in quasiparticle (QP) density throughout the array.
When these QPs tunnel across the qubits’ Josephson junctions, they induce correlated errors.
Engineering different superconducting gaps across the qubit’s Josephson junctions provides a method to resist this form of QP tunneling.
By fabricating all-aluminum transmon qubits with both strong and weak gap engineering on the same substrate, we observe starkly different responses during high-energy impact events.
Strongly gap engineered qubits do not show any degradation in $T_1$ during impact events, while weakly gap engineered qubits show events of correlated degradation in $T_1$.
We also show that strongly gap engineered qubits are robust to QP poisoning from increasing optical illumination intensity, whereas weakly gap engineered qubits display rapid degradation in coherence.
Based on these results, gap engineering mitigates the threat of high-energy impacts to QEC in superconducting qubit arrays.
\end{abstract}

%% file: content/1_introduction.tex
% \section{Introduction}\label{sec:introduction}

Scalable and useful quantum computing is achievable through the use of quantum error correction (QEC), provided the hardware satisfies key underlying assumptions.
One such assumption is the absence of any substantial source of correlated error.
In superconducting qubit arrays, this assumption is heavily violated by error bursts due to high-energy impacts~\cite{wilen_correlated_2021, cardani_reducing_2021, mcewen_resolving_2022, thorbeck_tls_2022, cardani_disentangling_2022, harrington_synchronous_2024, li_direct_2024}.
During an impact event, high-energy radiation deposits up to 1~MeV in the device substrate, which radiates throughout the device as high-energy phonons \cite{wilen_correlated_2021, kelsey_g4cmp_2023}.
As illustrated in \fig[a]{intro}, these phonons then break Cooper pairs, producing significantly elevated quasiparticle (QP) densities.
When these QPs tunnel through the Josephson junction, they can absorb the qubit excitation and induce excess $T_1$ decay~\cite{aumentado_nonequilibrium_2004, lutchyn_quasiparticle_2005, catelani_quasiparticle_2011, catelani_relaxation_2011, riste_millisecond_2012, wang_measurement_2014, gustavsson_suppressing_2016, serniak_hot_2018, christensen_anomalous_2019, serniak_direct_2019, diamond_distinguishing_2022}.
Impacts can therefore produce a massive correlated error burst that extends through time and space, which is not correctable using standard QEC.
The presence of these bursts induces an error floor on the logical performance of QEC experiments, preventing the promised exponential suppression of logical error with increasing code size~\cite{google_quantum_ai_exponential_2021, google_quantum_ai_suppressing_2022}.
Without mitigation in hardware, this effect single-handedly prevents scaling QEC to algorithmically relevant logical error rates.

\setlength{\textfloatsep}{0pt}
\setlength{\floatsep}{0pt}
\setlength{\intextsep}{0pt}
\begin{figure}[!hb]
    \vspace{0pt}
    \centering
    \resizebox{\linewidth}{!}{
        \includegraphics{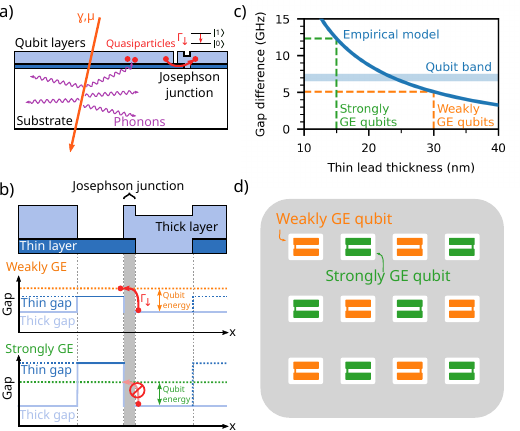}
    }
    \setlength{\abovecaptionskip}{-5pt}  
    \setlength{\belowcaptionskip}{0pt} 
    \caption{
    \textbf{Suppressing QP tunneling using gap engineering.} 
    a) High-energy impact events deposit large energies in the device substrate, which radiates as high-energy phonons.
    These phonons break Cooper pairs, generating QPs which tunnel through the Josephson junction and induce $T_1$ decay.
    b) QP tunneling by absorbing the qubit energy is only possible when the qubit energy exceeds the difference in superconducting gaps between the two junction leads.
    c) The superconducting gap of aluminum is sensitive to film thickness, permitting access to  different regimes of gap difference.
    d) The device consists of a checkerboard pattern of weakly (orange) and strongly (green) gap engineered transmon qubits. 
    A device diagram to scale is included in the Supplementary Material.
    }
    \label{fig:intro}
\end{figure}

While there have been several suggested~\cite{martinis_saving_2021} and implemented strategies for suppressing these effects, none have yet demonstrated total suppression of the error bursts.
Moving to a low-radiation underground facility~\cite{cardani_reducing_2021, gusenkova_operating_2022} greatly reduces the event rate of impacts, but remaining events such as those due to radioactive materials in the device and packaging remain unmitigated~\cite{cardani_disentangling_2022}.
Using phonon or QP traps~\cite{henriques_phonon_2019, iaia_phonon_2022, thorbeck_tls_2022, bargerbos_mitigation_2023} reduces both the temporal and spatial extent of the error burst, but QP poisoning is still observed.

Meanwhile, gap engineering has been shown to modify the dynamics of QP tunneling and improve the charge-parity lifetimes in superconducting qubits~\cite{aumentado_nonequilibrium_2004, lutchyn_quasiparticle_2005, court_quantitative_2008, gladchenko_superconducting_2009, bell_microwave_2012, sun_measurements_2012, riwar_efficient_2019, kalashnikov_bifluxon_2020, marchegiani_quasiparticles_2022, pan_engineering_2022, connolly_coexistence_2023, kamenov_suppression_2023}.
It has also been shown to improve the speed of recovery from QP poisoning from high-energy impact events~\cite{harrington_synchronous_2024}. 
Our approach to gap engineering involves producing different superconducting gaps on each side of the junction presents an energy barrier to QPs, as illustrated in \fig[b]{intro}.
The superconducting gap in thin films of aluminum is sensitive to thickness, with thinner films displaying higher superconducting gaps.
When the gap difference between the thin and thick leads of the Josephson junction ($\delta\Delta = \Delta_{\mathrm{thin}} - \Delta_{\mathrm{thick}}$) is significantly larger than the qubit energy ($E_q = h f_q$), QPs with energy near the gap can no longer absorb the qubit energy required to climb up the barrier to the higher gap material.
This suppresses the primary mechanism by which QPs induce $T_1$ decay.

Here, we investigate the prevalence of correlated error bursts on a single device with two types of transmon qubits: \textit{strongly gap engineered} (i.e. $\delta\Delta / h \gg f_q$) and \textit{weakly gap engineered} (i.e. $\delta\Delta / h \lesssim f_q$).
During impact events, the weakly gap engineered qubits display correlated energy relaxation errors.
Conversely, we observe no correlated $T_1$ decay errors on the strongly gap engineered qubits during impact events.
Further, we show that the strongly gap engineered qubits are robust to significant QP poisoning induced by optical illumination of the chip, enabling us to measure qubit coherence at controlled and elevated steady-state QP densities.
The robustness of this protection, coupled with the ease of achieving strong gap engineering, removes high-energy impact events as a roadblock from the path toward fault-tolerant quantum computing in superconducting qubits.

%% file: content/2_experiment.tex
\section{Experimental Details}\label{sec:experiment}

\subsection{Gap variation with thickness}

In aluminum, the superconducting gap $\Delta$ is sensitive to film thickness $t$.
While $\Delta$ in thick aluminum films ($t\gtrsim 100$~nm) is close to the bulk value ($\Delta \approx \Delta_{\mathrm{bulk}}$), for thinner films ($t\lesssim30$~nm) $\Delta$ grows quickly.
We use a simple empirical model~\cite{marchegiani_quasiparticles_2022} for the superconducting gap $\Delta=\Delta_{\mathrm{bulk}}+A/t$, where $\Delta_{\mathrm{bulk}}=180\,\mathrm{\mu eV}$ and $A=900\,\mathrm{\mu eV {\cdot} nm}$.
In \fig[c]{intro}, we use this model to predict the gap difference $\delta\Delta$ at the junction as a function of the thin lead thickness, holding the thick lead at 100~nm.
Designing qubits with different aluminum layer thicknesses on each side of the Josephson junction permits access to different regimes of gap engineering.

\subsection{Device Details}
We fabricate an array of 12 transmon qubits in an all-aluminum device.
We form a checkerboard placement of weakly and strongly gap engineered qubits to ensure that each class of qubit is uniformly distributed across the device, as illustrated in \fig[d]{intro}.
The qubits are frequency-tunable with maximum frequencies in the range $f_{\text{max}} \approx 7-7.3$~GHz and feature shared Purcell filters\cite{reed_fast_2010, jeffrey_fast_2014}.
We optimize the qubit geometry for coherence in this test device following Refs. ~\cite{place_new_2021, wang_measurement_2014}.
Each qubit footprint is approximately 0.6~mm $\times$ 0.8~mm, with junction leads that are 2~$\mu$m long and 150~nm wide.
We provide further information about the device fabrication and circuit parameters in the Supplementary Materials.

We achieve different levels of gap engineering by choosing the thicknesses of the aluminum layers.
Each qubit consists of two layers of aluminum with different thicknesses, which overlap in the capacitor pads and at the Josephson junction. 
The junction leads consist of only one layer of aluminum, one lead being thick and the other thin.
In all qubits, we choose the thick layer to be 100~nm to approach the bulk gap value $\Delta_{\mathrm{bulk}}$ and allow us to select over a wide range of gap differences $\delta\Delta$.
For ``weakly gap engineered'' qubits, we choose 30~nm for the thin layer.
This thickness targets a gap difference of around 5~GHz, substantially lower than the qubit energy for standard operating parameters.
This gap difference also allows us to flux-tune qubits to explore the regimes with qubit frequencies both above and below resonance with the gap difference.
For ``strongly gap engineered'' qubits, we choose 15~nm for the thin layer, which targets a gap difference of around 12~GHz, well above the maximum qubit frequency.

\subsection{Gap-engineering at the junction}
The thin junction lead provides a high-gap region separating the junction from the two regions of low-gap material - the thick junction lead on one side and the thick layer of the qubit capacitor pad on the other.
In a strongly gap engineered qubit, this barrier exceeds the energy scale of the qubit.
QPs near the superconducting gap in either capacitor pad then have insufficient energy to overcome the barrier and therefore either cannot reach or cannot tunnel across the junction.
As such, strongly gap engineered qubits will suppress the QP tunneling process that produces severe qubit decay, removing the dominant effect of elevated QP density on qubit $T_1$.
QPs should be mainly generated in the capacitor pads. 
QPs in the thin layer of the capacitor pads will fall into the lower-gap thick layer, rapidly depleting any stray QPs in the thin layer.
This should prevent any significant density of QPs moving from the thin layer of the qubit capacitor pad into the thin junction lead, so we do not expect to need additional depositions to achieve strong suppression of QP tunneling.

We note that the simplicity of our approach relies on the fact that the thin aluminum layer is the highest-gap material in our qubits. If the capacitor pad featured higher-gap material than the aluminum thin junction lead (e.g. Nb or Ta capacitor pads), the barrier would be compromised. QPs in the capacitor pad on the thin lead side would always be able to flow into the lead and across the junction. In this case, the barrier could instead be made using a yet-higher-gap material.

%% file: content/3_correlated_sampling.tex
\section{Rapid Repetitive Correlated Sampling}\label{sec:sampling}

\setlength{\textfloatsep}{18pt}
\setlength{\floatsep}{12pt}
\setlength{\intextsep}{12pt}
\begin{figure}[t]
    \centering
    \resizebox{\linewidth}{!}{
        \includegraphics{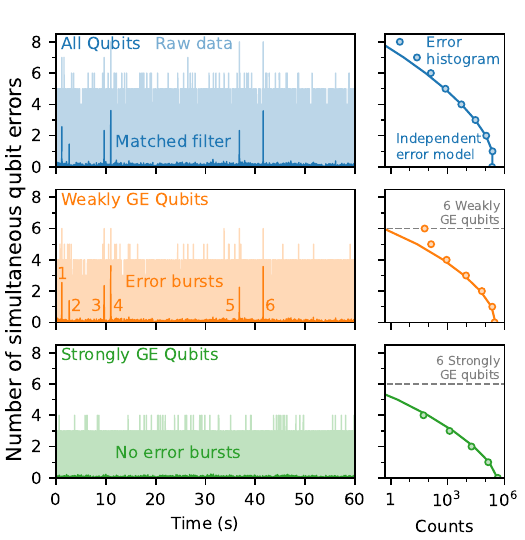}
    }
    \caption{
    \textbf{Correlated sampling time series.}
    Left: time series of simultaneous qubit errors summed over all qubits (top, blue), and different levels of gap engineering (GE): weakly gap engineered qubits (middle, orange) and strongly gap engineered qubits (bottom, green). 
    Both raw data (light) and matched filtered data are shown, illustrating the location of impact events (dark).
    Right: corresponding histograms over simultaneous qubit errors (light) and expected rates based on independent measurements of $T_1$ and readout fidelities (dark).
    The strongly gap engineered qubits behave according to the independent error model even during impact events.
    }
    \label{fig:sampling}
\end{figure}

We perform rapid repetitive correlated sampling as in Ref.~\cite{mcewen_resolving_2022}.
This involves initializing all qubits in the first excited state $\ket{1}$, idling for 1~\us and then measuring all qubits.
If the qubit is measured to be in the ground state $\ket{0}$, we consider that a qubit error. 
This cycle is repeated every 100~\us over the span of 60~seconds to form a single experiment of \num[group-separator={,}]{600000} samples.
The full dataset was produced by first independently measuring the T1 and readout fidelity on each qubit to inform the independent error model, followed by 100 correlated sampling experiments as described above. 
This procedure took around 8~hours total time and represents 6000~seconds total sampling time. 
As the typical timescale for variation in individual qubit T1 due to TLSs is only several hours~\cite{klimov_fluctuations_2018}, parameter drift can lead to small discrepancies between the independent model performance and the performance of qubits during the experiment. 
We find 154 high-energy impact events, corresponding to an average time between events of close to 40~seconds.
We include further analysis in the Supplementary Material.
The corresponding event flux ($2.6\mathrm{s}^{-1}\mathrm{cm}^{-2}$) matches well to the event flux found on larger devices ($2.1\mathrm{s}^{-1}\mathrm{cm}^{-2}$ \cite{mcewen_resolving_2022}).
As in Ref.~\cite{mcewen_resolving_2022}, when we perform the experiment initializing $\ket{0}$ considering we do not find any evidence of an excess rate of qubits being excited to $\ket{1}$ during impact events. 

\fig{sampling} shows a data from a single 60~second experiment featuring six events.
As in Ref.~\cite{mcewen_resolving_2022}, we use a matched filter to locate events with high confidence.
The matched filter exploits the reliable shape of the impact event in the error time-series, and is mathematically equivalent to a scaled cross-correlation with a template of the event shape.
Events are clearly visible on the weakly gap engineered qubits.
Correspondingly, the performance of these qubits deviates noticeably from the prediction from independent $T_1$ and readout errors.
For instance, all six weakly gap engineered qubits experience an error simultaneously two orders of magnitude more frequently than expected, indicative of highly correlated error bursts.
The same events do not affect the performance of the strongly gap engineered qubits, as they continue to match the independent error model.
In all 100 datasets, we find no time period where the behaviour of the strongly gap engineered qubits deviate from the independent error model, including where we see clear evidence of impacts in the weakly gap engineered qubits.

\begin{figure}[p!]
    \centering
    \resizebox{\linewidth}{!}{
        \includegraphics{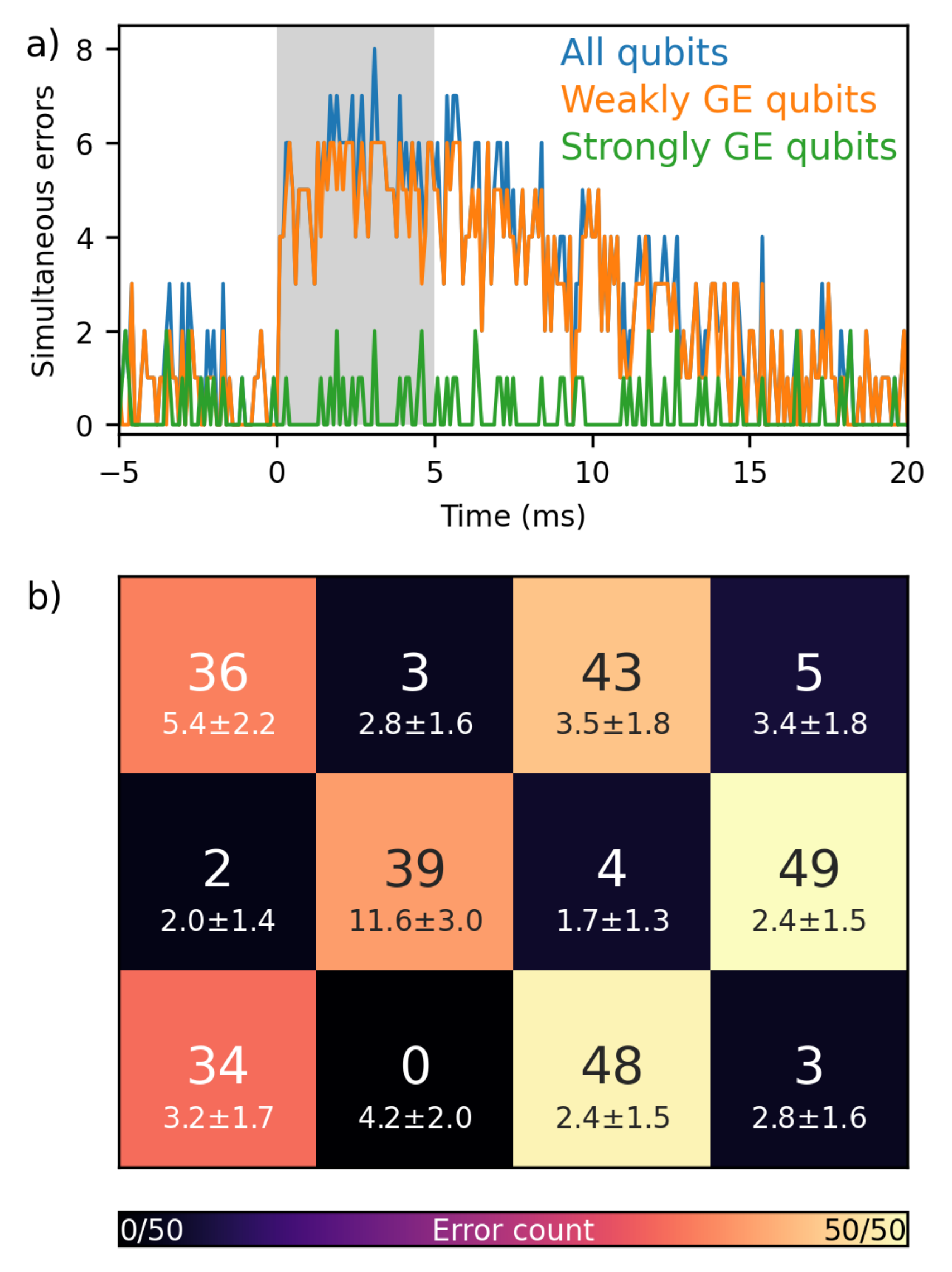}
    }
    \caption{
    \textbf{Errors during a single impact event.}
    a) Simultaneous qubit errors in 25~ms around an impact event, summed over all qubits (blue), and different levels of gap engineering (GE): weakly gap engineered qubits (blue) and strongly gap engineered qubits (green).
    b) Spatial distribution of errors during the 5~ms (50 measurement cycles) immediately following the impact event (grey highlight).
    We include the total number of errors and the expected number of error $\pm1\sigma$ from independent $T_1$ and readout errors.
    While the weakly gap engineered qubits experience greatly elevated error rates during the impact event, the strongly gap engineered qubits do not.
    }
    \label{fig:impact}
\end{figure}

\fig{impact} shows the time series data around a single impact event.
This is the event labeled $4$ shown in \fig{sampling}.
In \fig[a]{impact}, we see that the event affects only the weakly gap engineered qubits, elevating their error rate significantly.
By contrast, the strongly gap engineered qubits do not respond to the event, displaying error rates expected from independent $T_1$ and readout fidelities.
During the event, we see substantial variation in error, revealing the checkerboard pattern of strongly and weakly gap engineered qubits directly in \fig[b]{impact}.
During all discovered impact events, we see the SGE qubit display no additional error compared to their performance away from events and matching the independent error model.
This confirms that the SGE qubits are robust to QP-induced correlated errors from impact events.

%% file: content/4_illumination.tex
\begin{figure}[p]
    \centering
    \resizebox{\linewidth}{!}{
        \includegraphics{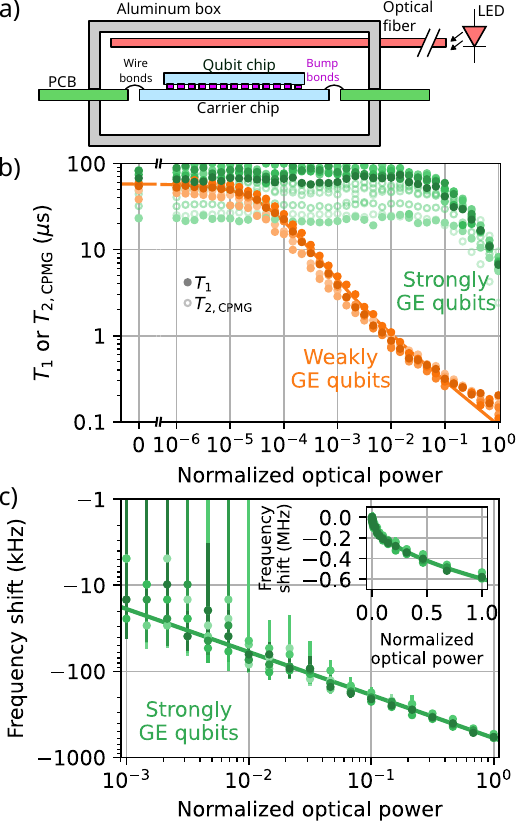}
    }
    \caption{
    \textbf{Measurement of qubits under illumination.}
    a) A sketch of the illumination setup.
    Light from a LED is coupled into the sample box using a fiber, producing flood illumination of the device.
    b) Qubit coherence under illumination.
    Weakly gap engineered qubits (orange circles; each qubit corresponds to a shade of orange) display a reduction in $T_1$ beginning at very low optical powers ($1 \times 10^{-4}$).
    Strongly gap engineered qubits display no significant change in $T_1$ (filled green circles; each qubit corresponds to a shade of green) or $T_{2,\mathrm{CPMG}}$ (open green circles) until high optical powers $1 \times 10^{-1}$.
    The weakly gap engineered qubits' response to optical power agrees with a steady-state QP density model (orange line).
    c) Frequency shifts of strongly gap engineered qubits under illumination.
    The solid line is a fit to a square-root dependence on optical power.
    Such a dependence is expected under a simple model where frequency shifts are proportional to QP densities and where steady-state QP density is limited by QP recombination.
    (Inset) Same data, plotted in linear scale.
    }
    \label{fig:illumination}
\end{figure}

\section{Illumination of the sample}\label{sec:light}

In order to further probe the qubit response to elevated QP density, we examine qubit coherence under the influence of optical illumination, which is used to introduce photons that break Cooper pairs and increase the QP density to a desired steady-state value.
Specifically, we consider the density of quasiparticles as a fraction of the density of cooper pairs $x_{\mathrm{qp}}$.
We use a light-emitting diode with a center wavelength of $\lambda = 1200$~nm whose output is fiber coupled into the sample box, as shown in \fig[a]{illumination}.
The silicon substrate is transmissive at these optical wavelengths, and allows for the light to penetrate through the qubit and carrier chips.
Unlike other works using optical illumination to implement QP injection~\cite{benevides_quasiparticle_2023}, our configuration employs continuous wave flood illumination of the entire device, allowing for approximately simultaneous and uniform application of optical power flux over all qubits. We are able to vary the optical illumination power by six orders of magnitude, as well as completely disabling it, where the full-scale normalized optical power of 1.0 corresponds to 16~$\mu\mathrm{W}$ measured outside of the dilution refrigerator.
Further details of the optical illumination setup can be found in the Supplementary Materials. 

\fig[b]{illumination} shows $T_1$ on all the qubits at their idle frequencies, which are in the range of 6.3~GHz to 6.6~GHz, in response to varying magnitudes of optical power.
As we increase the normalized optical power from $0$ to $10^{-3}$, the weakly gap engineered qubits (orange circles) suffer from large $T_1$ reductions of over an order of magnitude.
On the other hand, for the strongly gap engineered qubits (green circles), we find insignificant increase of the relaxation rates from QPs until the normalized optical power reaches $10^{-1}$.
We find similar robustness in CPMG dephasing lifetimes $T_{2,\mathrm{CPMG}}$ for the strongly gap engineered qubits.
The $T_2$ in the weakly gap engineered qubits is limited by the rapid reduction in qubit T1. 

In the strongly gap engineered qubits, the preservation of coherence at high optical illumination powers allows us to carry out measurements of qubit properties in the presence of high $x_{\mathrm{qp}}$.
One such measurement is the qubit frequency shift  from elevated $x_{\mathrm{qp}}$.
In order to account for slow qubit frequency drifts during the measurement, we use a differential Ramsey interferometry technique, where we measure the qubit frequency with no optical illumination interleaved with qubit frequency measurements at varying optical illumination powers.
The interleaving period is about 10 seconds and accounts for most quasistatic drifts.
\fig[c]{illumination} shows the qubit frequency shift $\Delta f_q$ of the strongly gap engineered qubits for normalized optical powers in the range of $10^{-3}$ to 1.
For this range of optical powers, we observe a square-root dependence of the qubit frequency shift on the optical power.
This is also the dependence between $x_{\rm qp}$ and the optical power since the magnitude of the fractional qubit frequency shift $\left|\Delta f_q/f_q\right|$ is proportional to $x_{\mathrm{qp}}$~\cite{catelani_quasiparticle_2011,catelani_relaxation_2011}.

For the weakly gap engineered qubits, we obtain the dependence of $x_{\rm qp}$ on the optical power from the qubit relaxation rates $T_1^{-1}(f_q)$, measured as a function of the qubit frequency and for normalized optical powers in the range of $3.2\times 10^{-5}$ to 1.
As shown in  Supplementary Materials, $x_{\rm qp}$ exhibits a linear dependence on the optical power until a normalized optical power of roughly $3\times10^{-4}$, and then $x_{\rm qp}$ exhibits a square-root dependence at larger optical powers.
Notably, the inferred $x_{\mathrm{qp}}$ from the fractional qubit frequency shifts of the strongly gap engineered qubits agrees well with the inferred $x_{\mathrm{qp}}$ from the $T_1^{-1}(f_q)$ spectrum of the weakly gap engineered qubits.
We estimate $x_{\mathrm{qp}} \simeq 2\times10^{-4}$ at the maximal optical illumination power, and $x_{\mathrm{qp}} \simeq 6\times10^{-5}$ is the point where the strongly gap engineered qubits begin to show degradation (corresponding to $10^{-1}$ normalized optical power in~\fig[b]{illumination}).
Given this density exceeds what is produced even in large impact events ($x_{\mathrm{qp}} \simeq 10^{-5}$~\cite{mcewen_resolving_2022}), this corroborates the insensitivity of strongly gap engineered qubits to the excess QP density induced by high-energy impact events.

%% file: content/5_conclusion.tex
\section{Summary and Conclusion}\label{sec:conclusion}

We have demonstrated a gap engineering technique that removes error bursts due to high-energy impact events.
Simply using a thinner aluminum junction lead, we protect our qubits from decoherence due to QP tunneling.
Unlike phonon or quasiparticle trapping, this requires no additional fabrication complexity, only requiring modified deposition of aluminum.

By suppressing the mechanism of qubit decay from QP tunneling rather than directly suppressing the QP density, this technique also opens new avenues for measurement of QP-induced effects.
For example, the qubit frequency shift at large QP densities was not previously resolvable given the strong suppression of $T_1$ in QP poisoned qubits.

That said, suppression of the QP population itself remains an important area of study.
The techniques of phonon and QP trapping will be important in reducing the remaining sources of loss, including QP scattering in the capacitor material.
We also note that gap engineering does not address the scrambling of two-level systems (TLSs) by high-energy impact events \cite{thorbeck_tls_2022}, which must be addressed either through improvements in the TLS density or via dynamic frequency-tuning to re-calibrate around the TLS scrambling in-situ.
As such, rather than replacing current mitigation strategies completely, our gap engineering technique represents a substantial advance in our first line of defense, being both effective and easy to implement.

Gap engineering at the junction opens the way to strong protection from the effects of QP poisoning in large-scale devices.
Future QEC experiments protected from the error floor due to impact events are now free to demonstrate the key building blocks of fault-tolerant quantum computing at scale.

%% file: content/6_declarations.tex
\section*{Contributions}
Matt McEwen, Kevin Miao, Alex Opremcak designed the experiment.
Juan Atalaya, Kevin Miao provided modeling of the qubit response to QPs.
{Alex Opremcak}, Alex Crook, Jenna Bovaird, John~Mark Kreikebaum contributed to the design and fabrication of the experimental device.
Kevin Miao, Alex Opremcak, Alex Bilmes, Nick Zobrist, Evan Jeffrey constructed the experimental apparatus.
Kevin Miao, Matt McEwen, Alex Bilmes calibrated and measured the experimental device.
Bicheng Ying, Matt McEwen contributed key software infrastructure for performing measurements.
Alex Bilmes, Andreas Bengtsson, Alex Crook, Hung-Shen Chang, Andrew Dunsworth conducted measurements of test devices.
Julian Kelly, Yaxing Zhang provided modeling of limiting loss mechanisms.
Ebrahim Forati provided electromagnetic simulations of devices.
Rajeev Acharya, Justin Iveland, Wayne Liu, Seon Kim, and Brian Burkett contributed to the fabrication, characterization, and processing techniques used in the devices.
Matt McEwen, Kevin Miao, Alex Opremcak compiled the manuscript.
All authors contributed to revising the manuscript.

\section*{Acknowledgements}
We thank Lev Ioffe and Lara Faoro for helpful discussions.
We thank Alexander Korotkov for helpful comments on the manuscript.
We thank the entire Google Quantum AI team for maintaining the hardware, software, cryogenics, and electronics infrastructure that enabled this experiment, and for producing an environment where this work is possible. 

%% file: content/S1_device.tex
\section{Device Description}\label{sec:supp_device}

\begin{figure}[h!]
    \centering
    \resizebox{\linewidth}{!}{
        \includegraphics{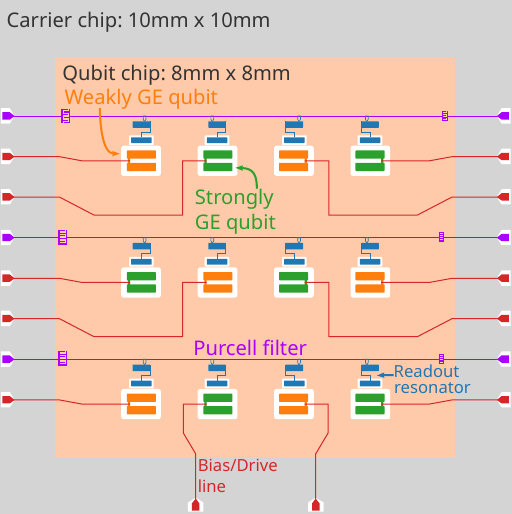}
    }
    \caption{
    \textbf{A sketch of the device layout to scale.} 
    The qubit chip features six weakly gap engineered qubits (orange) and six strongly gap engineered qubits (green) and a surrounding ground-plane (beige). 
    This chip is hybridized to a larger carrier chip (grey) which includes readout resonators (blue), shared Purcell filters (purple) and combined qubit bias and drive lines (red).
    }
    \label{fig:device}
\end{figure}

Our device consists of two separate high-resistivity silicon chips, hybridized together via bump bonds as depicted in \fig{device}.
Aluminum is used as the superconducting metal for all features on both chips, apart from the bump bonds. 
The carrier chip is approximately 10mm $\times$ 10mm in area with a thickness of 525 $\mu$m, and contains the qubit control wiring, readout resonators, Purcell filters, and ground plane \cite{reed_fast_2010, jeffrey_fast_2014}.
The qubit chip is approximately 8mm $\times$ 8mm in area with a thickness of 380 $\mu$m, and consists of 12 frequency-tunable transmon qubits and ground plane.
As discussed in the main text, the transmon qubits are of two types: (i) 6 ``strongly gap engineered'' qubits with Josephson junction electrode thicknesses of 15nm/100nm \& (ii) 6 ``weakly gap engineered'' qubits with Josephson junction electrode thicknesses of 30nm/100nm (see \tbl{ge_by_qubit}).
We fabricate the two sets of qubits in sequential lithography steps, first the weakly gap-engineered qubits and the ground plane, then the strongly gap-engineered qubits in a second step.
Both types of qubits are uniformly distributed across the qubit chip to avoid any position-dependent biasing.
The ground planes on the carrier and qubit chips are galvanically connected by the bump bonds.

Parameters for the qubit system are listed in \tbl{qubit_anharmonicities} -\tbl{qubit_t1s} according to their physical arrangement on the device (see \tbl{ge_by_qubit} and Fig. 1(d).) 
We note that the idle qubit frequencies and idle qubit $T_1$'s were changed by small amounts throughout our experiments to combat fluctuating two-level system defects \cite{klimov_fluctuations_2018}.
Parameters for the qubit-resonator system are listed in \tbl{resonator_frequencies} -\tbl{resonator_fids}.

\begin{table}[h!]
\caption{\label{tbl:ge_by_qubit} Gap Engineering Style by Qubit.}
    \begin{tabular}{ | c  | p{1.25cm}  | p{1.25cm} | p{1.25cm} | p{1.25cm} |}
    \hline
    & \multicolumn{4}{|c|}{Qubit Column} \\ \cline{1-5}
    \multirow{3}{*}{\makecell{Qubit\\Row}} & \hfil weak & \hfil strong & \hfil weak & \hfil strong \\ \cline{2-5}
    & \hfil strong & \hfil weak & \hfil strong & \hfil weak \\ \cline{2-5}
    & \hfil weak & \hfil strong & \hfil weak & \hfil strong \\ \hline
    \end{tabular}
\end{table}

\begin{table}[h!]
\caption{\label{tbl:qubit_anharmonicities} Qubit Anharmonicities (MHz).}
    \begin{tabular}{ | c  | p{1.25cm}  | p{1.25cm} | p{1.25cm} | p{1.25cm} |}
    \hline
    & \multicolumn{4}{|c|}{Qubit Column} \\ \cline{1-5}
    \multirow{3}{*}{\makecell{Qubit\\Row}} & \hfil 198 & \hfil 196 & \hfil 198 & \hfil 200 \\ \cline{2-5}
    & \hfil 206 & \hfil 197 & \hfil 199 & \hfil 198 \\ \cline{2-5}
    & \hfil 198 & \hfil 199 & \hfil 198 & \hfil 199 \\ \hline
    \end{tabular}
\end{table}

\begin{table}[h!]
\caption{\label{tbl:qubit_idles} Idle Qubit Frequencies (GHz).}
    \begin{tabular}{ | c  | p{1.25cm}  | p{1.25cm} | p{1.25cm} | p{1.25cm} |}
    \hline
    & \multicolumn{4}{|c|}{Qubit Column} \\ \cline{1-5}
    \multirow{3}{*}{\makecell{Qubit\\Row}} & \hfil 6.480 & \hfil 6.320 & \hfil 6.370 & \hfil 6.410 \\ \cline{2-5}
    & \hfil 6.510 & \hfil 6.490 & \hfil 6.400  & \hfil 6.450 \\ \cline{2-5}
    & \hfil 6.460 & \hfil 6.385 & \hfil 6.420 & \hfil 6.550 \\ \hline
    \end{tabular}
\end{table}

\begin{table}[h!]
\caption{\label{tbl:qubit_f_maxes} Maximum Qubit Frequencies (GHz).}
    \begin{tabular}{ | c  | p{1.25cm}  | p{1.25cm} | p{1.25cm} | p{1.25cm} |}
    \hline
    & \multicolumn{4}{|c|}{Qubit Column} \\ \cline{1-5}
    \multirow{3}{*}{\makecell{Qubit\\Row}} & \hfil 7.01 & \hfil 7.29 & \hfil 6.95 & \hfil 7.28 \\ \cline{2-5}
    & \hfil 7.30 & \hfil 7.08 & \hfil 7.33 & \hfil 7.04 \\ \cline{2-5}
    & \hfil 7.14 & \hfil7.31 & \hfil7.11 & \hfil 7.25 \\ \hline
    \end{tabular}
\end{table}

\begin{table}[h!]
\caption{\label{tbl:qubit_t1s} Idle Qubit $T_1$ ($\mu$s).}
    \begin{tabular}{ | c  | p{1.25cm}  | p{1.25cm} | p{1.25cm} | p{1.25cm} |}
    \hline
    & \multicolumn{4}{|c|}{Qubit Column} \\ \cline{1-5}
    \multirow{3}{*}{\makecell{Qubit\\Row}} & \hfil 53 & \hfil 23 & \hfil 51 & \hfil 63 \\ \cline{2-5}
    & \hfil 82 & \hfil 49 &  \hfil 61 & \hfil 50 \\ \cline{2-5}
    & \hfil 62 & \hfil 55 & \hfil 52 & \hfil 58 \\ \hline
    \end{tabular}
\end{table}

\begin{table}[h!]
\caption{\label{tbl:resonator_frequencies} Resonator Frequencies (GHz).}
    \begin{tabular}{ | c  | p{1.25cm}  | p{1.25cm} | p{1.25cm} | p{1.25cm} |}
    \hline
    & \multicolumn{4}{|c|}{Qubit Column} \\ \cline{1-5}
    \multirow{3}{*}{\makecell{Qubit\\Row}} & \hfil 7.43 & \hfil 7.38 & \hfil 7.34 & \hfil 7.33 \\ \cline{2-5}
    & \hfil 7.41 & \hfil 7.36 & \hfil 7.32 & \hfil 7.31 \\ \cline{2-5}
    & \hfil 7.43 & \hfil 7.38 & \hfil 7.34 & \hfil 7.33 \\ \hline
    \end{tabular}
\end{table}

\begin{table}[h!]
\caption{\label{tbl:resonator_kappas} Resonator Decay Times (ns).}
    \begin{tabular}{ | c  | p{1.25cm}  | p{1.25cm} | p{1.25cm} | p{1.25cm} |}
    \hline
    & \multicolumn{4}{|c|}{Qubit Column} \\ \cline{1-5}
    \multirow{3}{*}{\makecell{Qubit\\Row}} & \hfil 35 & \hfil 16 & \hfil 8 & \hfil 43 \\ \cline{2-5}
    & \hfil 33 & \hfil 22 & \hfil 23 & \hfil 53 \\ \cline{2-5}
    & \hfil 31 & \hfil 16 &  \hfil7 & \hfil 36 \\ \hline
    \end{tabular}
\end{table}

\begin{table}[h!]
\caption{\label{tbl:resonator_qubit_couplings} Resonator-Qubit Couplings (MHz).}
    \begin{tabular}{ | c  | p{1.25cm}  | p{1.25cm} | p{1.25cm} | p{1.25cm} |}
    \hline
    & \multicolumn{4}{|c|}{Qubit Column} \\ \cline{1-5}
    \multirow{3}{*}{\makecell{Qubit\\Row}} & \hfil 68 & \hfil 59 & \hfil 69 & \hfil 45 \\ \cline{2-5}
    & \hfil 61 & \hfil 65 & \hfil 52 & \hfil 60 \\ \cline{2-5}
    & \hfil 60 & \hfil 73 & \hfil 58 & \hfil 54 \\ \hline
    \end{tabular}
\end{table}

\begin{table}[h!]
\caption{\label{tbl:resonator_fids} Readout Assignment Fidelity (\%).}
    \begin{tabular}{ | c  | p{1.25cm}  | p{1.25cm} | p{1.25cm} | p{1.25cm} |}
    \hline
    & \multicolumn{4}{|c|}{Qubit Column} \\ \cline{1-5}
    \multirow{3}{*}{\makecell{Qubit\\Row}} & \hfil 98 & \hfil 99 & \hfil 97 & \hfil 97 \\ \cline{2-5}
    & \hfil 99 & \hfil 96 & \hfil 98 & \hfil 95 \\ \cline{2-5}
    & \hfil 98 & \hfil 99 & \hfil 97 & \hfil 98 \\ \hline
    \end{tabular}
\end{table}

%% file: content/S2_rrecs_series.tex
\section{Statistics of impact events}\label{sec:supp_series}

\begin{figure}[b!]
    \centering
    \resizebox{\linewidth}{!}{
        \includegraphics{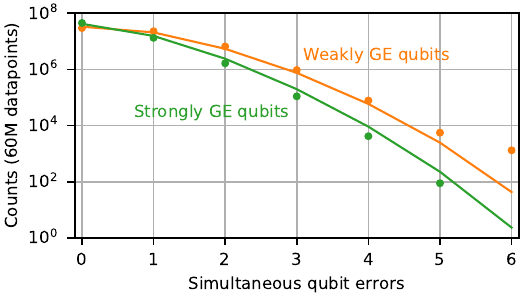}
    }
    \caption{
    \textbf{Histogram of simultaneous errors over all datasets.} 
    Counts (circles) of simultaneous errors over all 100 datasets (60 million total datapoints) for weakly gap engineered (orange) and strongly gap engineered (green) qubits.
    We also include predicted error counts from independent $T_1$ and readout fidelity measurements (solid lines).
    The weakly gap engineered qubits display a characteristic excess of higher numbers of simultaneous errors than predicted, indicative of impact events.
    The strongly gap engineered qubits do not show any excess errors over the independent prediction.
    }
    \label{fig:histogram}
\end{figure}

We took 100 Rapid Repetitive Correlated Sampling (RReCS)~\cite{mcewen_resolving_2022} datasets of 60~seconds long each, with a 1~\us sampling time between state initialization and measurement, and a 100~\us sampling interval between measurements.
\fig{histogram} shows the histogram of simultaneous error counts summed over all datasets.
The weakly gap engineered qubits show rates of simultaneous errors that are substantially higher than the independent prediction from measured T1 and readout errors, while the strongly gap engineered qubits do not.

To locate events, we used the same matched filter analysis technique as in Ref.~\cite{mcewen_resolving_2022}. 

We exploit the reliable shape of events in the time series to locate them by using a matched filter. 
We use a template function
\[ 
\mathrm{Template(t)} =
\begin{cases}
    a\exp(-(t-t_0)/\tau_{\mathrm{decay}}) + c &; t \geq t_0\\
    c &; t < t_0
\end{cases}
\]
to generate the filter, using parameters $\tau_{\mathrm{decay}}=8$~ms, $a=1$, $c=0$, $t_0=0$. Applying the filter is mathematically equivalent to removing the DC component from the time-series and then convolving the timeseries by the time-reversed template function.
Our selection of $\tau_{\mathrm{decay}}=8$~ms compared to $\tau_{\mathrm{decay}}=20$~ms in Ref.~\cite{mcewen_resolving_2022} wll only influence the symmetry and scale of the peaks in the matched-filtered time-series. 
Because of the dependence of the matched-filtered time-series on the choice of template, we only use it for identifying peak locations, and return to the raw data to fit for event magnitude and recovery time. 

We split the data by qubit type, with 0 events being found for the strongly gap engineered qubits.
For the weakly gap engineered qubits, we located 154 peaks in total over the 6000~seconds of data, giving an average event rate of 1 per 38.96~seconds for a 10~mm $\times$ 10~mm chip area.

\begin{figure}[b!]
    \centering
    \resizebox{\linewidth}{!}{
        \includegraphics{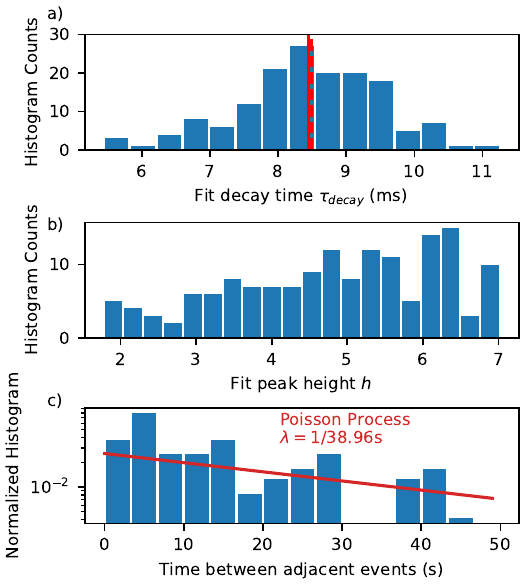}
    }
    \caption{
    \textbf{Parameter distributions for impact events.} 
    Parameter distributions fit from 154 impact events using only the weakly gap engineered qubit data.
    a) Fit exponential decay constant $\tau_{\mathrm{decay}}$.
    We also indicate the mean (solid) and median (dashed).
    b) Fit absolute peak heights.
    These heights occasionally exceed 6 due to inaccuracy of the exponential fit template, which does not account for finite rise time or the non-exponential decay in real event traces.
    c) A normalized histogram of the time elapsed between the 75 neighbouring pairs of events in the same dataset.
    We include the expected values for a Poisson process with $\lambda$ equal to the average event rate over all the data (1 per 38.96~seconds).
    This is included for comparison to Ref.~\cite{mcewen_resolving_2022}, but the lower average event rage on this smaller device and the length of individual datasets (60~seconds) greatly limits the usefulness of this analysis.
    }
    \label{fig:series}
\end{figure}

\fig{series} shows the fitted parameter distributions over the 154 located events.
We find that the characteristic time for recovery is around 8.5~ms.
This is substantially shorter than the $\sim$25~ms in Ref.~\cite{mcewen_resolving_2022}, but still substantially longer than other values reported in the literature in the 1 to 100~\us regimes \cite{wilen_correlated_2021, thorbeck_tls_2022, iaia_phonon_2022}.

%% file: content/S3_illumination.tex
\section{Optical illumination setup}\label{sec:supp_light}

In order to introduce excess QP densities to the sample in a controlled fashion, we use optical light with large enough photon energies ($E_{\gamma} \sim 1~\mathrm{eV}$) to break Cooper pairs in the superconducting aluminum ($2\Delta \sim 200~\mu\mathrm{eV}$) throughout the device.
In order to minimally disturb the microwave environment of the qubit sample box, we opted to use a light source external to the dilution refrigerator and transmit the light to the sample through an optical fiber.
While this prevents us from easily focusing or rastering the light across the sample, it ensures good thermal and microwave properties of the qubits measured in this study, and minimizes unwanted noise sources that may further degrade qubit coherence when the light intensity is modulated.
The optical fiber enters the qubit sample box via a small machined hole and terminates to free space near a corner of the box.
The emitted light then reflects on the walls of the aluminum sample box and flood illuminates the sample within the box.

Our choice of the excitation source (Thorlabs; M1200F1) is determined by a suitable emission wavelength that is transmissive through silicon, while also transmissive through common, off-the-shelf optical fibers.
With a center wavelength of 1200~nm, the excitation is not significantly attenuated by the silicon substrates of the qubit and carrier chips, while also considerably reflected by the aluminum sample box and deposited wiring.
The attenuation through the optical fiber (Thorlabs; FG105UCA) at 1200~nm is expected to be of order 1 $\mathrm{dB}/\mathrm{m}$.
This allows for near-complete coverage of optical excitation across the entire device with appreciably strong optical powers.
We control the intensity of the light using a LED driver (Thorlabs; LEDD1B) modulated by an external voltage controller.
We calibrate the optical power emitted by the LED using an InGaAs photodiode power sensor (Thorlabs; S154C).
We estimate an attenuation through optical components and optical fiber from the output of the LED source to the sample box of 36 dB.
At the intermediate stages within the dilution refrigerator, we coil the fiber into small loops and thermalize those loops to the stages in order to eject far-infrared photons coupled into the fiber from hotter stages.

%% file: content/S4_qp_model.tex
\section{Modeling quasiparticle properties under illumination}\label{sec:qp_model}

\begin{figure*}[ht!]
    \centering
    \includegraphics[width=\textwidth]{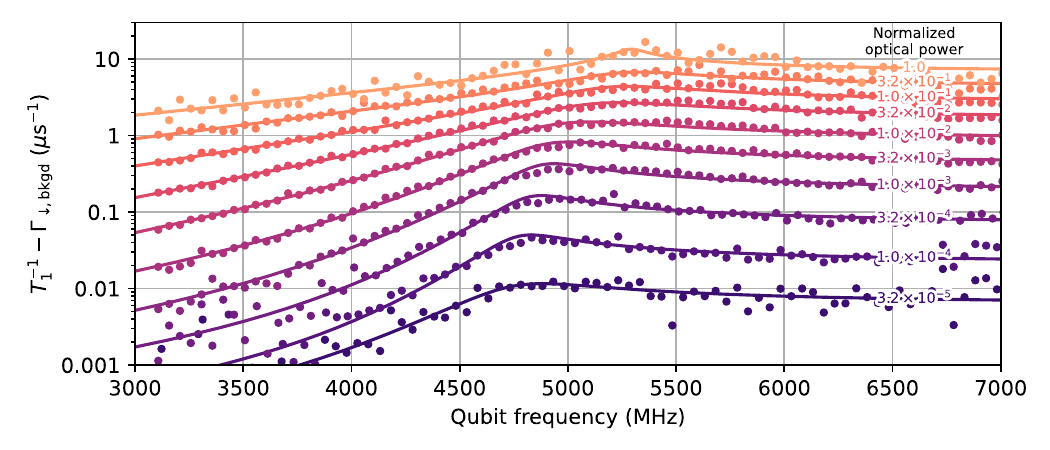}
    \caption{
    \textbf{Light-induced qubit decay from quasiparticle tunneling.}
    $1/T_1$ spectrum data (colored circles) for one of the weakly gap engineered qubits at varying powers of optical illumination.
    The background decay contribution $\Gamma_{\downarrow,\mathrm{bkgd}}$ is subtracted from each $1/T_1$ spectrum.
    The spectra are then individually fitted to the model described by \eq{gamma_qp}.
    The peaks in each spectrum around 5000~MHz correspond to the gap difference of the Josephson junction leads for this particular qubit.
    }
    \label{fig:T1_qp_fit}
\end{figure*}

\begin{figure*}[ht!]
    \centering
    \includegraphics[width=\textwidth]{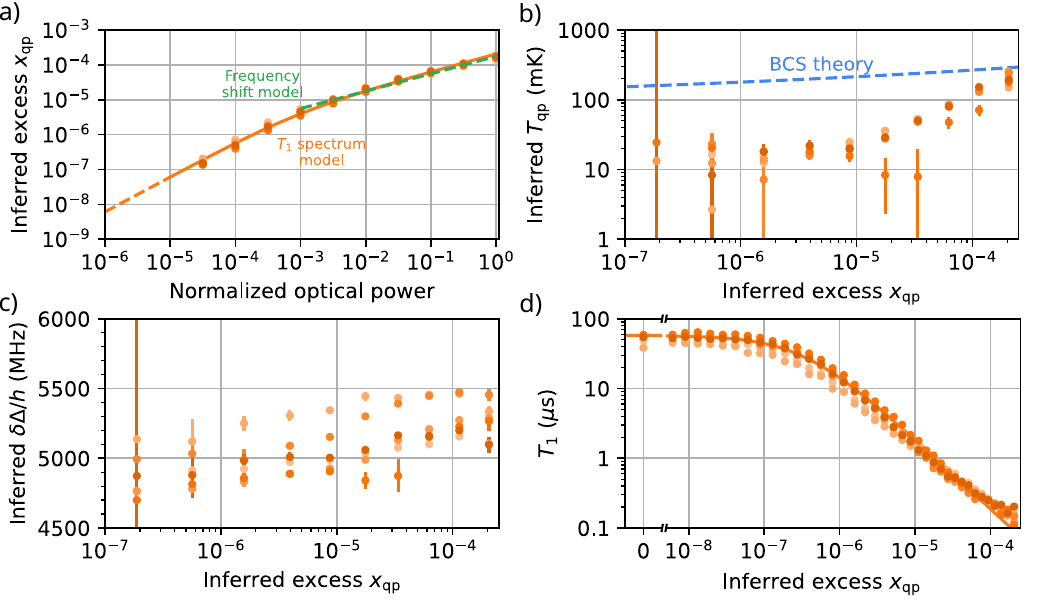}
    \caption{
    \textbf{Modeled quasiparticle properties under illumination.} 
    a) Inferred excess QP density $x_{\rm qp}$  from the $1/T_1$ spectrum model (\eq{gamma_qp}) of the weakly gap engineered qubits (orange circles; each qubit corresponds to a shade of orange).
    The average of the weakly gap engineered qubits' $x_{\rm qp}$ is fit to the steady-state $x_{\rm qp}$ formula (\eq{steady_state_xqp}; solid orange line)Using this formula, we extrapolate below resolvable differences in $1/T_1$ down to a normalized optical power of $10^{-6}$ (dashed orange line).
    The average $x_{\rm qp}$ of the strongly gap engineered qubits is obtained from the qubit frequency shifts shown in~Fig.~4(c) and~\eq{relative_freq_shift_formula} with $a=0.5$ (dashed green line).
    The overlap of the solid orange and dashed green lines suggests that the amount of optically-induced $x_{\rm qp}$ is similar in both types of gap engineered qubits.
    b) Effective quasiparticle temperature $T_\mathrm{qp}$ extracted from the $1/T_1$ spectrum model of the weakly gap engineered qubits.
    BCS theory predicts quasiparticle densities comparable to those measured in this work at significantly elevated temperatures (dashed blue line).
    c) Inferred gap difference $\delta \Delta$ of the weakly gap engineered qubits as a function of $x_\mathrm{qp}$.
    d) $T_1$ data of the six weakly gap engineered qubits presented in Fig.~4(b) with the horizontal axis remapped to excess $x_\mathrm{qp}$.
    }
    \label{fig:qp_model}
\end{figure*}

Here, we explain how to extract QP densities $x_\mathrm{qp}$ introduced by optical illumination at various optical powers using two different methods, one for each of the two kinds of qubits.

For weakly gap engineered qubits, we obtain $x_{\rm qp}$ from the dependence of the qubit decay rate $1/T_1$ on the qubit frequency (we call this dependence $1/T_1$ spectrum), see~\fig{T1_qp_fit}.
The qubit decay rate is given by $$1/T_1 = \Gamma_{\downarrow,\mathrm{bkgd}} + \Gamma_{\downarrow, \mathrm{qp}},$$ where $\Gamma_{\downarrow,\mathrm{bkgd}}$ is the (background) qubit decay rate in the absence of light and $\Gamma_{\downarrow, \mathrm{qp}}$ denotes the additional contribution to $1/T_1$ due to QPs generated from optical illumination that tunnel through the Josephson junctions.
This additional contribution $\Gamma_{\downarrow, \mathrm{qp}}$ is proportional to the sought-after QP density $x_{\rm qp}$ as follows~\cite{catelani_relaxation_2011}:
\begin{widetext}
\begin{align}
     \Gamma_{\downarrow, \mathrm{qp}} (f_q) = 2 f_q x_\mathrm{qp} \displaystyle\int\limits_{\Delta_{\rm thick}}^{\infty} {\rm d}\varepsilon\, \frac{1}{\sqrt{\varepsilon - \Delta_{\rm thick}}} \operatorname{Re}{\left({\frac{1}{\sqrt{\varepsilon + hf_q - \Delta_{\rm thin} + ihw }}}\right)} \exp{\left({-\frac{\varepsilon - \Delta_{\rm thick}}{k_{\rm B}T_\mathrm{qp}}}\right)} \sqrt{\frac{\Delta_{\rm thick}}{2\pi k_{\rm B}T_\mathrm{qp}}}. 
    \label{eq:gamma_qp}
\end{align}
\end{widetext}
In \eq{gamma_qp}, $\Delta_{\rm thick}$ and $\Delta_{\rm thin}$ are the superconducting energy gaps of the thick and thin leads, so that $\Delta_{\rm thick} < \Delta_{\rm thin}$.
The expression for $\Gamma_{\downarrow,\mathrm{qp}}\left(f_q\right)$ accounts for the contribution to the qubit decay due to absorption of the qubit energy $hf_q$ during tunneling of QPs from the lower superconducting gap lead into the higher superconducting gap lead (the contribution to the qubit decay due to tunneling in the opposite direction is neglected; we assume that the population of QPs in the thin lead is negligible compared to the QP population in the thick lead).
The broadening parameter $w$ in \eq{gamma_qp} is a phenomenological parameter introduced to avoid the  divergence of $\Gamma_{\downarrow,\mathrm{qp}}$ when $h f_q = \Delta_{\rm thin} - \Delta_{\rm thick}$. 

\fig{T1_qp_fit} shows the fit (solid colored lines) of the $1/T_1$ spectrum using \eq{gamma_qp} for one of the weakly gap engineered qubits at various light intensities.
Other weakly gap engineered qubits have similar goodness of fit and are not shown here.
% The fit works well and similar fits (not shown) are obtained for the $T_1$ data of the other weakly gap engineered qubits.
The fit to the model involves four fitting parameters: the gap difference $\delta \Delta \equiv \Delta_{\rm thin} - \Delta_{\rm thick}$, the effective QP temperature $T_{\rm qp}$, the QP density $x_{\rm qp}$, and $w$. The $w$ parameter is only important to reproduce the observed ``roundness'' of the $1/T_1(f_q)$ curve near its peak.

In \fig[a]{qp_model}, we show the dependence of the inferred QP density $x_{\rm qp}$ on the normalized optical power (a normalized optical power of 1 corresponds to $16~\mu$W).
At low optical power, we find that $x_{\rm qp}$ exhibits a linear dependence, which crosses over to a square-root dependence at higher optical illumination powers (see discussion below).
\fig[b]{qp_model} shows the inferred QP temperature $T_{\rm qp}$ as a function of the inferred QP density $x_{\rm qp}$. For low optical powers, $T_{\rm qp}$ is around 20~mK and then significantly rises when the inferred QP density reaches $x_{\rm qp}\gtrsim 10^{-5}$.
The blue dashed line in \fig[b]{qp_model} indicates the temperature $T_{\rm BCS}$ at which the equilibrium QP density ($x_{\rm qp}^{\rm eq.} = \sqrt{2\pi k_{\rm B}T_{\rm BCS}/\Delta}\, \exp(-\Delta/k_{\rm B}T_{\rm BCS})$, with $\Delta/h = 50$~GHz) predicted by BCS would reach the inferred QP density $x_{\rm qp}$. Note that, as we increase the optical illumination power, the inferred QP temperature $T_{\rm qp}$ approaches the BCS temperature $T_{\rm BCS}$, suggesting some degree of heating of the silicon substrate and/or aluminium leads by the optical illumination.
\fig[c]{qp_model} shows the inferred superconducting gap difference $\delta \Delta/h \approx 5$~GHz which, in contrast to $x_{\rm qp}$ and $T_{\rm qp}$, only weakly depends on the optical illumination power.
The broadening parameter $w$ is of order $150$~MHz.
Lastly, in~\fig[d]{qp_model}, we replot the $T_1$ data of~Fig. 4(b) for the weakly gap engineered qubits as a function of $x_\mathrm{qp}$, using the monotonic dependence of $x_{\rm qp}$ on normalized optical power shown in~\fig[a]{qp_model}.

For strongly gap engineered qubits, we obtain $x_{\rm qp}$ from the relative qubit frequency shifts under varying optical illumination powers, using the formula~\cite{catelani_quasiparticle_2011,catelani_relaxation_2011}
\begin{align}
    \frac{\Delta f_q}{f_q} = -a\, x_\mathrm{qp},
    \label{eq:relative_freq_shift_formula}
\end{align}
where the proportionality coefficient is equal to $a=(2\pi)^{-1}\big[\sqrt{2\Delta/(\delta\Delta+hf_q)} + \sqrt{2\Delta/(\delta\Delta-hf_q)}\big]/2$, assuming $\delta\Delta > h f_q$. For a gap difference of $\delta\Delta/h = 12$~GHz and qubit frequency $f_{q}=6.5$~GHz, we obtain $a\approx 0.52$ (we use $\Delta/h = $50~GHz).
For the data shown in~Fig. 4(c), we find that the inferred $x_{\rm qp}$ exhibits a square-root dependence on the optical illumination power, for normalized optical powers larger than $10^{-3}$. \
This dependence is depicted by the green dashed line labeled ``Frequency shift model'' in~\fig[a]{qp_model}, using $a=0.5$.
Notably, we observe good agreement for the extracted excess $x_{\rm qp}$ from measuring qubit frequency shifts on the strongly gap engineered qubits and from independently measuring the $1/T_1$ spectra on the weakly gap engineered qubits.
This demonstrates our ability to uniformly introduce QPs across the device and among the two types of gap engineered qubits.

Finally, we discuss the dependence of the inferred QP density $x_{\rm qp}$ on optical illumination power. As shown in \fig[a]{qp_model}, the dependence is linear for low optical powers and becomes square-root-like at high optical powers.
This can be understood using the following model for the dynamics of QP density,
\begin{align}
    \dot{x}_\mathrm{qp}(t) = -s {x}_\mathrm{qp} - r{x}^2_\mathrm{qp} + g,
    \label{eq:x_qp_dynamics}
\end{align}
where $s$ is the single-particle trapping rate, $r$ is the recombination rate, and $g$ is the QP generation rate that includes a contribution proportional to the normalized optical power.
Note that $s$ and $r$ have the same physical units since $x_{\rm qp}$ is dimensionless.
In the steady-state regime we set the LHS of~\eq{x_qp_dynamics} to zero and find the stationary value of $x_{\mathrm{qp}}$, 
\begin{align}
    x_\mathrm{qp} = \frac{2g}{s + \sqrt{s^2 + 4gr}}.
    \label{eq:steady_state_xqp}
\end{align}
At low optical illumination powers ($g\lesssim s^2/4r$), the QP generation process is balanced by the trapping process and we obtain $x_{\rm qp} \simeq g/s$, which is proportional to the normalized optical power if QP generation is still dominated by the optical light source.
On the other hand, at high optical illumination powers, the QP generation process is balanced by the recombination process and we obtain $x_{\rm qp} \simeq \sqrt{g/r}$, which scales as the square-root of the normalized optical power.
This explains the change of scaling of $x_{\rm qp}$ with the normalized optical power seen in~\fig[a]{qp_model}. Fitting the inferred $x_{\rm qp}$ shown in~\fig[a]{qp_model} with~\eq{steady_state_xqp} leads to a ratio of recombination to trapping rates equal to $r/s = 1.4\times10^{5}$. Recombination rates are of the order $r \sim 10^7~{\rm s}^{-1}$~\cite{mcewen_resolving_2022, wang_measurement_2014}, so we obtain a trapping rate of order $s\sim 700~{\rm s}^{-1}$.
This value lies in the span of trapping rates found in the literature, ranging from 10~s$^{-1}$~\cite{diamond_distinguishing_2022} to $10^3$~s$^{-1}$~\cite{wang_measurement_2014}.